\newif\ifSC
\ifSC
\documentclass[onecolumn,draftclsnofoot,12pt]{IEEEtran}
\else
\documentclass[twocolumn,10pt]{IEEEtran}
\fi
\usepackage[utf8]{inputenc}
\usepackage{amsmath,amssymb,amsthm}
\usepackage{verbatim,bbm}
\usepackage{graphicx}
\usepackage{color}
\usepackage{framed}
\usepackage{epstopdf}
\usepackage{soul}
\usepackage{etoolbox}
\IEEEoverridecommandlockouts

\newtoggle{SC}
\ifSC
\usepackage{setspace}
\settoggle{SC}{true}
\else
\settoggle{SC}{false}
\fi

\newcommand{\dplus}{{3\rm{D}^{+}}}

\newcommand{\oldJeffC}[1]{}
\newcommand{\newAdd}[1]{#1}
\newcommand{\newRem}[1]{}

\ifSC
\newcommand{\scf}{.48}
\newcommand{\rsgo}{.48}
\newcommand{\rsgt}{.60}
\else
\newcommand{\scf}{.48}
\newcommand{\rsgo}{.40}
\newcommand{\rsgt}{.43}
\fi
\newcommand{\Eq}{}

\newcommand{\prob}[1]{ \mathbb{P}\left[#1\right] }
\newcommand{\expect}[1]{\mathbb{E}\left[#1\right] }

\newcommand{\laplace}[1]{\mathcal{L}_{#1} }

\newcommand{\Ball}[0]{ \mathcal{B} }

\newcommand{\SINR}{\text{SINR}}

\newcommand{\ie}{{\em i.e. } }

\newcommand{\xinchenpaper}{ZhangAndrews2015}

\newcommand{\Pc}{\mathrm{P_c}}

\newcommand{\Pcu}[2]{\mathrm{P}_\mathrm{#1}^{#2}}
\newcommand{\Cfunc}[2]{C\left(#1,#2\right)}
\newcommand{\Ifunc}[4]{I\left(#3,#4\right)}

\newcommand{\intd}{\text{d}}
\newcommand{\z}{\text{z}}

\newtheorem{lemma}{Lemma}
\newtheorem{theorem}{Theorem}
\theoremstyle{definition}
\newtheorem{definition}{Definition}

\title{SINR and Throughput Scaling in Ultradense Urban Cellular Networks}
\author{Abhishek K. Gupta, Xinchen Zhang, Jeffrey G. Andrews
\thanks{A. K. Gupta ({\tt g.kr.abhishek@utexas.edu}) and J. G. Andrews ({\tt jandrews@ece.utexas.edu}) are with Wireless Networking and Communications Group, Department of Electrical and Computer Engineering at the University of Texas at Austin, Austin, TX 78712 USA. Xinchen Zhang is with Qualcomm Inc.}}

\begin{document}
\maketitle
\begin{abstract}
We consider a dense urban cellular network where the base stations (BSs) are stacked vertically as well as extending infinitely in the horizontal plane, resulting in a greater than two dimensional (2D) deployment. Using a dual-slope path loss model that is well supported empirically, we extend recent 2D coverage probability and potential throughput results to 3 dimensions. We prove that the ``critical close-in path loss exponent'' $\alpha_0$ where SINR eventually decays to zero is equal to the dimensionality $d$, i.e. $\alpha_0 \leq 3$ results in an eventual SINR of 0 in a 3D network.  We also show that the potential (i.e. best case) aggregate throughput decays to zero for $\alpha_0 < d/2$. Both of these scaling results also hold for the more realistic case that we term $\dplus$, where there are no BSs below the user, as in a dense urban network with the user on or near the ground. 
\end{abstract}

\def \podname {PaperContent}

\bstctlcite{IEEEexample:BSTcontrol}

\section{Introduction}
Cellular networks have been continually densifying their base stations (BSs) since their inception, driving most of the increased throughput over the past several decades \cite{Andrews5G}.  The urban environments where spectrum is most scarce and high capacity most critical are themselves continually densifying, particularly vertically, albeit at a slower rate.  Future cellular deployments, featuring small low power base stations, will extend in the vertical direction as well as in the 2D plane.  Recently,  \cite{\xinchenpaper} showed that for a 2D network, the coverage probability for a given SINR value decays to zero as the network is heavily densified if the close-in (i.e. within a distance $R_c$) path loss exponent $\alpha_0$ is less than 2, regardless of the other path loss exponents outside $R_c$.  The potential throughput, i.e. the best case aggregate throughput, still grows at-least sub-linearly if $\alpha_0 > 1$.  

\textit{Other Related Work:}  The coverage probability of cellular wireless systems has been studied in detail in recent years using stochastic geometry \cite{Andrews2011,Bai2015} for various 2D deployment scenarios. In \cite{Brown2014}, a general $d$ dimensional Poisson Point Process (PPP) BS deployment is considered and an equivalent one dimensional (1D) PPP is derived to compute the SINR and rate coverage for highest instantaneous power based association.  The dual-slope path loss model \cite{Sarkar2003} is a generalization of standard single slope path loss model, which was analyzed in \cite{\xinchenpaper}  and is the focus of this letter. It is well supported by many measurements \cite{Erceg1992,Xia1994,Hampton2006} and shown to be very close to many scenarios of current interest including indoor \cite{Has93}, LTE \cite{3GPP,WINNERII} and millimeter wave  \cite{Rap2013,Chang2014}.

\textit{Contributions:}
The contribution of this letter is to extend \cite{\xinchenpaper} to three dimensions. The typical 3D case corresponds to a user sufficiently high off the ground in a dense urban environment that they see an appreciable number of BSs in every direction.   We also consider a case we term $\dplus$, which is a special case of 3D with BSs extending overhead in the positive direction only, corresponding more closely to a user on the ground. We compute the probability of coverage (SINR distribution) and the potential throughput for both of these cases. Then we compute the critical values of the close-in path loss exponent for which SINR and throughput respectively go to zero as the density goes to infinity for both the 3D and $\dplus$ cases.

\section{System Model}
We consider a downlink cellular network  with  BSs located in a $d$-dimensional space according to a Poisson Point Process (PPP) $\Phi=\{x_i : x_i \in \mathbb{R}^d\}$ with intensity $\lambda$.  The average number of BSs in a $d$-ball of radius $r$ located at origin for such a PPP is given by 
\iftoggle{SC}{ $\Lambda(\Ball(r))=V_d r^d \lambda$ }{
\begin{align*}
\Lambda(\Ball(r))=V_d r^d \lambda
\end{align*}}
where $V_dr^d$ is the volume of $d$-ball $\Ball(r)$ and $V_d$ is a constant dependent on the dimension $d$. Note that $V_{2}=\pi$ and $V_{3}=\frac43 \pi$, while for the $\dplus$ case, BSs are located only in a half sphere, so $V_{\dplus}= V_{3}/2 = \frac23\pi$. With a slight abuse of notation, we will use $d$ = 3D and $\dplus$ to differentiate between 3D and $\dplus$ cases where necessary, keeping in mind that $d=3$ for both cases.

We assume a dual-slope path loss model which has two different path loss exponents, as in \cite{\xinchenpaper}, given as
\begin{align*}
\ell(r)&=
\begin{cases}
r^{-\alpha_0}, &\text{for } r\le R_c\\
\eta r^{-\alpha_1}, &\text{for } r\ge R_c
\end{cases},
\end{align*}
where $R_c$ is the critical distance, $\alpha_0$ is the close-in path loss exponent and $\alpha_1$ is the long-range path loss exponent, with $\eta = R_c^{\alpha_1-\alpha_0}$ a constant to provide continuity. We require $0\le \alpha_0\le\alpha_1$, and assume  Rayleigh fading for all links, therefore the received power at the origin from the $i^{th}$ BS located at $x_i$ is given by $P_i=h_i \ell(\|x_i\|)$, where $h_i$'s are i.i.d. exponential random variables with mean 1.

We assume that the user connects to the BS providing the highest average received power ({i.e.} closest BS) and denote this BS by index $0$. Therefore the SINR is 
\begin{align*}
\SINR&=\frac{h_0 \ell(\|x_0\|)}
	{\sigma^2+\sum_{i\in \Phi\setminus \{0\}}{h_i \ell(\|x_i\|)}}
\end{align*}
where $\sigma^2$ is the noise variance.

We are interested in the following two performance metrics. 
\begin{definition}
The downlink \emph{coverage probability} in $d$ dimensions is
\begin{align*}
\Pcu{SINR}{d}(\lambda,T)= \prob{\SINR>T},
\end{align*}
which is equivalently the ccdf of the SINR.
\end{definition}

\begin{definition}
The \emph{potential throughput}  $\tau^d_l$ captures the average number of bits that can be transmitted per unit area per unit time per unit bandwidth, assuming all BSs transmit (i.e. full buffer model), and is
\begin{align*}
\tau^d_l(\lambda,T)=\log_2(1+T) \lambda \Pcu{SINR}{d}(\lambda,T).
\end{align*}
It has units of area spectral efficiency: $\mathrm{bps/Hz/m^2}$.
\end{definition}

\section{Probability of Coverage}
In this section, we first derive the coverage probability expression for general path loss function and present the simplified expression for the dual-slope path loss model.

\begin{lemma}
\label{lemma1}
The coverage probability\newRem{ for the} with a general path loss function is 
\ifSC
\begin{align}
&\Pcu{SINR}{d}(\lambda,T)=\lambda V_d \int_0^\infty  e^{-T\sigma^2/ \ell(y^{\frac1d})}  \times 
\exp{\left(
-\lambda V_d y 
	\left(
		1+\int_1^\infty \frac{T}{T+\frac{\ell(y^{\frac1d})}{\ell((ty)^{\frac1d})}}\intd t
	\right)
\right)} \intd y. \label{Eq:PcExp}
\end{align}
\else
\begin{align}
&\Pcu{SINR}{d}(\lambda,T)=\lambda V_d \int_0^\infty  e^{-T\sigma^2/ \ell(y^{\frac1d})} \nonumber \\
& \times 
\exp{\left(
-\lambda V_d y 
	\left(
		1+\int_1^\infty \frac{T}{T+\frac{\ell(y^{\frac1d})}{\ell((ty)^{\frac1d})}}\intd t
	\right)
\right)} \intd y. \label{Eq:PcExp}
\end{align}
\fi
For the \rm{3D} and $\dplus$ cases, $V_d$ is  $\frac{4\pi}{3} $ and $\frac{2\pi}{3}$, respectively.
\end{lemma}

\begin{figure}
 \centering
 \includegraphics[width=\scf\textwidth]{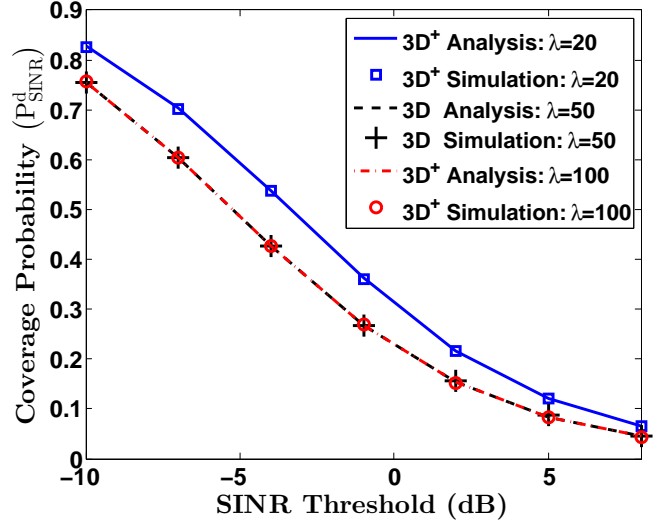}
 \caption{SINR coverage probability for $\dplus$ and 3D BS deployment with $\alpha_0=3.3, \alpha_1=5, \sigma^2=1, R_c=0.4$. }
 \label{fig:25D3DsimAnalComp}
 \end{figure}

\begin{IEEEproof}See Appendix \ref{proofoflemma1}.
\end{IEEEproof}
We observe that expression for 3D and $\dplus$ are the same, except for the value of\newRem{$d$ and thus} $V_d$.  Also, the effective BS density for $\dplus$ is $\frac{\lambda}{2}$ when compared to the 3D case.  Thus, from now on, we will consider only the 3D case for analysis with the understanding that such results can be trivially converted to $\dplus$.
 Fig. \ref{fig:25D3DsimAnalComp} validates Lemma \ref{lemma1} by comparing a simulation of the system model with the analytic expression given in the Lemma. It also shows that the SINR coverage of a $\dplus$ deployment with density $\lambda$ is equal to that of a 3D deployment with density $\frac{\lambda}{2}$.
 
Lemma \ref{lemma1} can be further simplified for the dual-slope case 
to give the  following Theorem.

\begin{theorem} \label{theorem:SINR}
The downlink coverage probability for a general $d$-dimensional PPP BS deployment under the dual-slope model is given as \iftoggle{SC}{$\Pcu{SINR}{d}(\lambda,T)=$}{}\begin{align}
\iftoggle{SC}{}{\Pcu{SINR}{d}(\lambda,T)=}\lambda V_d R_c^d \int_0^1
	e^{
		-\lambda V_d R_c^d \Ifunc{\alpha_0}{\alpha_1}{T}{r}-T\sigma^2R_c^{\alpha_0}r^{\frac{\alpha_0}{d}} 
	}
	\intd r \iftoggle{SC}{}{ \nonumber\\ }
	+\lambda V_d R_c^d \int_1^\infty 
	e^{
		-\lambda V_d R_c^d \Cfunc{-\frac{\alpha_1}{d}}{T}r-T\sigma^2R_c^{\alpha_0} r^{\frac{\alpha_1}{d}} 
	}
	\intd r \label{eq:theorem1}
\end{align}  
where\iftoggle{SC}{$\Ifunc{\frac{\alpha_0}{d}}{\alpha_1}{T}{r}$ $ =$ $\Cfunc{\frac{\alpha_0}{d}}{\frac{1}{Tr^{\frac{\alpha_0}{d}}}}    $ $
+$  $ \Cfunc{-\frac{\alpha_1}{d}}{Tr^{\frac{\alpha_0}{d}}}$ $
-r\Cfunc{\frac{\alpha_0}{d}}{\frac1T}$ $ +$ $r-1$ and \newline $\Cfunc{b}{z}=$
  $ \ _2F_1\left(1,\frac1b,1+\frac1b,-z\right),$
}{\begin{align*}
\Ifunc{\frac{\alpha_0}{d}}{\alpha_1}{T}{r}=&\Cfunc{\frac{\alpha_0}{d}}{\frac{1}{Tr^{\frac{\alpha_0}{d}}}}    
+\Cfunc{-\frac{\alpha_1}{d}}{Tr^{\frac{\alpha_0}{d}}}
\iftoggle{SC}{}{\\&}
-r\Cfunc{\frac{\alpha_0}{d}}{\frac1T}+r-1,
\\ \Cfunc{b}{z}=&\ _2F_1\left(1,\frac1b,1+\frac1b,-z\right),
\end{align*}}
with $\ _2F_1\left(a,b,c,z\right)$ being the Gauss hypergeometric function.
\end{theorem}
\begin{IEEEproof} See Appendix \ref{proofoftheorem1}.
\end{IEEEproof}
 Before going further, we will also compute the SIR coverage probability assuming noise to be zero  which mimics the interference limited case. SIR coverage probability tightly upper bounds SINR coverage probability\newRem{closely} for dense deployments and is 
%
 given as 
\begin{align*}
\iftoggle{SC}{}{ & }
\Pcu{SIR}{d}(\lambda,T)
\ifSC  \else \\ \fi
&=\lambda V_d R_c^d \left( \int_0^1
	e^{
		-\lambda V_d R_c^d \Ifunc{\alpha_0}{\alpha_1}{T}{r}
	}
	\intd r 
	+
	\int_1^\infty 
	e^{
		-\lambda V_d R_c^d \Cfunc{-\frac{\alpha_1}{d}}{T}{r}
	}
	\intd r\right)\\ 
	&=\lambda V_d R_c^d \int_0^1
	e^{
		-\lambda V_d R_c^d \Ifunc{\alpha_0}{\alpha_1}{T}{r}
	}
	\intd r 
	+
	\frac{e^{
		-\lambda V_d R_c^d \Cfunc{-\frac{\alpha_1}{d}}{T}
		}}
	{
	\Cfunc{-\frac{\alpha_1}{d}}{T}}.
\end{align*}

The SNR coverage probability can be found similarly by letting the interference go to zero.

%
%

The following Lemma establishes the relationship between the 2D case considered in \cite{\xinchenpaper} and the general $d$-dimensional case.

\begin{lemma}\label{lemma:equivalence}
The probability of ({\rm SIR, SINR} and {\rm SNR}) coverage  for a general $d$ dimension PPP BS deployment with parameters $\alpha_0,\alpha_1,\lambda,R_c,\sigma^2$ is equal to the probability of coverage for a \rm{2D} system with $\alpha_0^\prime , \alpha_1^\prime,\lambda^\prime,R_c,{\left(\sigma^{\prime}\right)}^2$ if
\begin{align*}
\alpha_0^\prime&=\frac2d\alpha_0, &
\alpha_1^\prime&=\frac2d\alpha_1, \iftoggle{SC}{&}{\\ }
\lambda^\prime&=\frac{R_c^d}{R_c^2}\frac{V_d}{V_2}\lambda ,&
{\left(\sigma^{\prime}\right)}^2&=\sigma^2R_c^{\alpha_0-\alpha_0^\prime}.
\end{align*}
\end{lemma}
\begin{IEEEproof}
Proven easily by substituting the respective parameters in \Eq \eqref{eq:theorem1} for $d=2$\newRem{ and $d=3$} and observing the exact same expression\newRem{for the two cases}.\oldJeffC{XXX Technically, this only proves it for the 2D vs. 3D case.  What about 2.5D or other dimensionalities?  Similarly for below Figure on equivalence you might at least then want to show it for 2.5D as well, if we decide to leave that figure in.}
\end{IEEEproof}



Following the similarity of SINR expression to that in \cite{\xinchenpaper}, it can be shown that 
\cite[Lemma 2]{\xinchenpaper} and \cite[Theorem 2]{\xinchenpaper} will also be valid for the general $d$ dimensional case. Building on these results and Lemma \ref{lemma:equivalence}, we state the following Theorem.

\begin{theorem}
\label{lemma:goestozero}
Under the dual-slope path-loss model, the SIR and SINR coverage probability of a general $d$-dimensional system go to 0 as $\lambda\rightarrow\infty$ for $\alpha_0\le d$.
\end{theorem}
\begin{IEEEproof}
See Appendix \ref{proofofgoestozero}.
\end{IEEEproof}

The above Theorem is true for  general $d$ dimensional deployments and hence is valid for both the $\dplus$ and 3D cases. It provides the critical values of the close-in path loss exponent below which the coverage probability goes to zero.  Theorem implies that for both the 3D and $\dplus$ scenario, the critical value of $\alpha_0$ is $3$.\oldJeffC{XXX This is not correct is it???  Should be 2.5 for 2.5D.  I'm very confused.}  It is very common for the path loss exponent of short range systems to be less than these $\alpha_0$ values, so this is seemingly an important concern for future ultra dense networks.

\begin{figure}
\centering
\includegraphics[width=\scf\textwidth,trim= 0 0 0 20,clip]{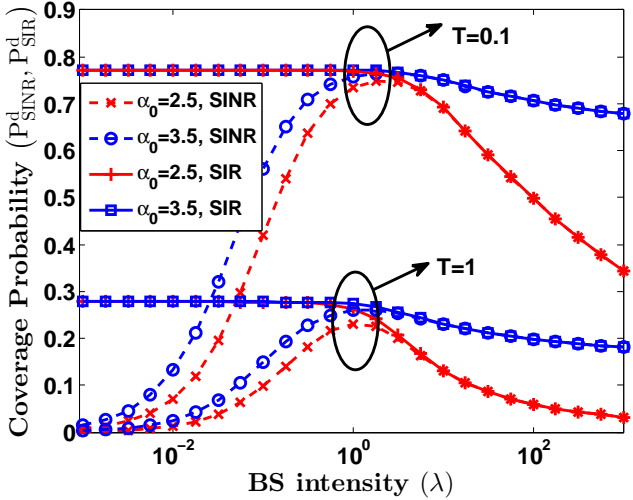}
\caption{SINR and SIR coverage scaling vs. network density ($\lambda$) for 3D deployment, with $\alpha_0 = [2.5,3.5]$, $\alpha_1 = 4$, $R_c = 0.4, \sigma^2 = 1$.}
\label{fig:SINRSIRcov}
\end{figure}

Fig. \ref{fig:SINRSIRcov} shows the behavior of SINR and \newAdd{SIR} coverage probability ($\Pcu{SINR}{3D}$ and $\Pcu{SIR}{3D}$) for a 3D BS deployment as the network density varies. It can be observed that for all path loss exponents, $\Pcu{SINR}{3D}$ first increases as $\lambda$ increases. After a critical limit of $\lambda$, $\Pcu{SINR}{3D}$ starts decreasing. For $\alpha_0$ less than 3, $\Pcu{SINR}{3D}$ goes to zero while for $\alpha_0<3$,  $\Pcu{SINR}{3D}$ asymptotically becomes a nonzero constant as  $\lambda\rightarrow \infty$.\newRem{ Similarly Fig. 4 shows the behavior of SIR coverage  probability $\Pcu{SINR}{3D}$ for a 3D BS deployment as the network density varies.} For lower $\lambda$, $\Pcu{SIR}{3D}$ corresponds to coverage probability for single slope path loss model with  $\alpha_1$ and for higher $\lambda$, $\Pcu{SIR}{3D}$ corresponds to that with $\alpha_0$. We can observe that $\Pcu{SIR}{3D}$ goes to zero for $\alpha_0<3$ as $\lambda$ goes to infinity.

\begin{figure}
\centering
\includegraphics[width=\scf\textwidth,trim=0 0 0 2,clip]{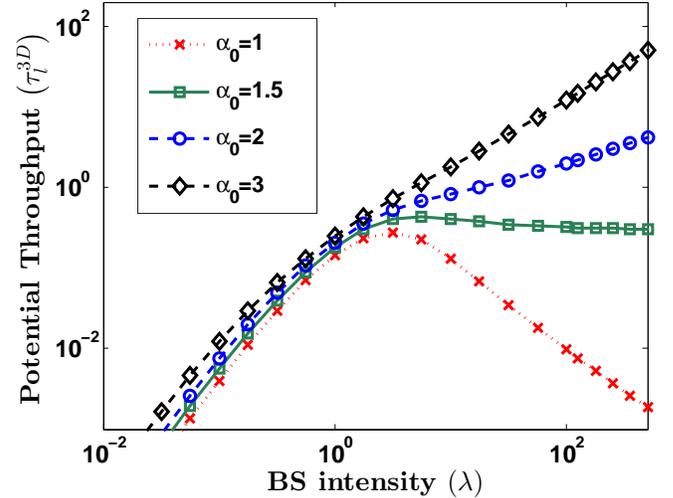}
\caption{Potential throughput ($\tau_l$) scaling with network density for 3D deployment. Here, $\alpha_0 =[1,1.5,2,3]$, $\alpha_1 = 4$, $R_c = 0.4, \sigma^2 = 1,T=1$.}
\label{fig:throughputscaling}
\end{figure}

\section{Potential Throughput}
We now turn to the potential throughput scaling with density. 
\begin{theorem}\label{theorem:throughputscaling}
Under the dual-slope model, as $\lambda\rightarrow \infty$, the potential throughput  $\tau_l^d$
\begin{enumerate}
\item  grows linearly with $\lambda$ if $\alpha_0>d$,
\item grows sublinearly with rate $\lambda^{(2-\frac{d}{\alpha_0})}$ if $\frac{d}{2}<\alpha<d$,
\item decays to zero if $\alpha_0<\frac{d}2$.
\end{enumerate}
\end{theorem}
\begin{IEEEproof}
Using Lemma \ref{lemma:equivalence}, we can prove that the potential throughout in a general $d$ dimensional  BS deployment is connected to that of 2D case by the following relation:
\begin{align*}
\tau_l^d(\lambda,T,\alpha_0,\alpha_1,R_c,\sigma^2)&=\frac{R_c^2}{R_c^d}\frac{V_2}{V_d} \tau^{2D}_l(\lambda^\prime,T,\alpha^\prime_0,\alpha^\prime_1,R_c,{\left(\sigma^\prime\right)}^{2}).
\end{align*}
Using the above relation and \cite[Theorem 3]{\xinchenpaper}, all three results of the Theorem \ref{theorem:throughputscaling} can be easily proven.  For $\alpha_0 > d$, \cite[Theorem 3]{\xinchenpaper}  states  that potential throughput in  2D case scales linearly with $\lambda^\prime$ for $\alpha^\prime_0>2$. Therefore, the potential throughput in the general $d$-dimensional case  will scale linearly with $\lambda$ if $\alpha=\alpha^\prime \frac{d}{2} >d$. Similarly, the other two results for $\frac{d}{2}<\alpha_0<d$ and $\alpha_0<\frac{d}{2}$ can be obtained.
\end{IEEEproof}

Fig. \ref{fig:throughputscaling} shows the scaling of potential throughput with respect to $\lambda$ for a 3D deployment. As expected, for $1.5<\alpha_0<3$ the potential throughput scales only sub-linearly. Theorem \ref{theorem:throughputscaling} provides a theoretical basis for understanding the gain in throughput vs. the cost of densification.

We conclude by noting that the SINR throughput and SINR scaling results can be easily extended to more than two path loss exponents. Owing to the equivalency between the 2D case discussed in \cite{\xinchenpaper} and the general $d$ dimensional case, Theorem \ref{lemma:goestozero} and Theorem \ref{theorem:throughputscaling} can be shown to be true for the multi-slope path loss model also.  

\appendices

\section{Proof of Lemma \ref{lemma1}} \label{proofoflemma1}
Let us denote the sum interference at origin  by $I$
. Now the  SINR coverage probability can be written as
\begin{align}
\Pcu{SINR}{d}&=\prob{\SINR>T}=\prob{\frac{h_0 \ell(\|x_0\|)}
	{\sigma^2+I}>T}
	\iftoggle{SC}{}{\nonumber\\
&}=\int_0^\infty e^{-T\sigma^2/\ell(x)} \laplace{I}(T/\ell(x))f_{\|x_0\|}(x)\intd x \label{appAe1}
\end{align}
where $f_{\|x_0\|}(x)$ is the probability distribution of the distance $\|x_0\|$ of the closest (serving) BS from origin given as $f_{\|x_0\|}(x)=V_d d\lambda x^{d-1}e^{-\lambda V_d x^d}$
and $\laplace{I}(s)$ is the Laplace transform of interference which  can be derived as 
\begin{align*}
& \laplace{I}(s)=\expect{e^{-sI}}=\expect{e^{-s\sum_{\z\in \Phi, \|\z\|>x}{h_\z \ell(\|\z\|)}}} \iftoggle{SC}{}{\\
&}=\exp\left( -\lambda \int_x^\infty \frac{s\ell(z)}{1+s\ell(z)} V_d d z^{d-1}\intd z\right).
\end{align*}
Substituting the values of $f_{\|x_0\|}(x)$ and $\laplace{I}(s)$ in \Eq \eqref{appAe1}, we get 
\begin{align*}
\Pcu{SINR}{d}&=V_d d\lambda\int_0^\infty e^{-T\sigma^2/\ell(x)}
 x^{d-1}e^{-\lambda V_d x^d}
\iftoggle{SC}{}{ \\ & \ \ }
\exp\left( -\lambda \int_x^\infty \frac{\ell(z)T/\ell(x)}{1+\ell(z)T/\ell(x)} V_d d z^{d-1}\intd z\right)
\intd x\\
&\stackrel{(a)}{=} V_d d\lambda\int_0^\infty e^{-T\sigma^2/\ell(x)}
 x^{d-1}e^{-\lambda V_d x^d}
\iftoggle{SC}{}{ \\ & \ \  }
\exp\left(- \lambda  V_d d x^{d}\int_1^\infty \frac{T}{\ell(x)/\ell(ux)+T} u^{d-1} \intd u\right)
\intd x
\end{align*}
where (a) is due to the substitution $z/x \rightarrow u$. Now using the substitutions $x^d\rightarrow y$ and $u^d\rightarrow t$, we get \Eq \eqref{Eq:PcExp}.

\section{Proof of Theorem \ref{theorem:SINR}} \label{proofoftheorem1}

Using the substitution, $r=y/R_c^d$ in \Eq  \eqref{Eq:PcExp}, we get 
\begin{align}
&\Pcu{SINR}{d}(\lambda,T)
=\lambda V_d R_c^d\int_0^\infty  e^{-T\sigma^2/ \ell(R_cr^{\frac1d})} 
\iftoggle{SC}{}{ \nonumber \\ & \times }
\resizebox{\rsgo\textwidth}{!}{$\displaystyle
\exp{\left(
-\lambda V_d R_c^d r
	\left(
		1+\int_1^\infty \frac{T}{T+\frac{\ell(R_cr^{\frac1d})}{\ell(R_c(tr)^{\frac1d})}}\intd t
	\right)
\right)} \intd r $}\label{AppBe1}.
\end{align}

Outer integration from $r=0$ to $\infty$ in \Eq \eqref{AppBe1} can be divided into integration over the following two intervals:
\begin{enumerate}
\item \textit{In interval $[0,1)$:} the inner integral can be written as
\begin{align}
&\int_1^{\frac{1}{r}} \frac{T}{T+{t^{\frac{\alpha_0}{d}}}}\intd t
		+\int_{\frac{1}{r}}^\infty \frac{T}{T+ t^{\frac{\alpha_0}{d}} {r^{\frac{\alpha_1-\alpha_0}{d}}}}\intd t \nonumber\\
& 
\ \ =1+\frac{1}{r}\Cfunc{\frac{\alpha_0}{d}}{\frac{1}{Tr^{\frac{\alpha_0}{d}}}}-\Cfunc{\frac{\alpha_0}{d}}{\frac{1}{T}}
\iftoggle{SC}{}{ \nonumber\\
&\ \ \ \ \ 
}
 +\frac1r\Cfunc{-\frac{\alpha_1}{d}}{-Tr^{\frac{\alpha_1}{d}}r^{-\frac{\alpha_1-\alpha_0}{d}}}-\frac1r
.\label{eq:intp1}
 \end{align}
 \item \textit{In interval $[1,\infty)$:} the inner integral can be written as
 \begin{align}
&\int_1^{\infty} \frac{T}{T+{t^{\frac{\alpha_1}{d}}}}\intd t
 =
 \Cfunc{-\frac{\alpha_1}{d}}{T}-1
 . \label{eq:intp2}
 \end{align}
 \end{enumerate}
 
 Using \eqref{eq:intp1} and \eqref{eq:intp2} in \Eq \eqref{AppBe1}, we get \Eq \eqref{eq:theorem1}.


\section{Proof of Theorem \ref{lemma:goestozero}} \label{proofofgoestozero}
This proof is similar to the proof of Proposition 1 in \cite{\xinchenpaper}. 
As the SINR coverage probability is always less than SIR coverage probability, it suffices to show the proof for SIR only. Using Lemma 1 and taking $\sigma^2=0$, we can upper bound the SIR coverage probability as following: \iftoggle{SC}{}{$\Pcu{SIR}{d}(\lambda,T)$}
\begin{align*}
\iftoggle{SC}{\Pcu{SIR}{d}(\lambda,T)}{}&\le 
\resizebox{\rsgt\textwidth}{!}{$\displaystyle
\lambda V_d \int_0^\infty 
\exp{\left(
-\lambda V_d y 
	\left(
		1+\int_1^{\max{(1,\frac{R_c^d}{y})}}\frac{T}{T+\frac{\ell(y^{\frac1d})}{\ell((ty)^{\frac1d})}}\intd t
	\right)
\right)} $}\intd y
\iftoggle{SC}{}{ \\
&\ \ \ }+ 
\lambda V_d \int_0^{R_c^d}  
e^{
-\lambda V_d y } \intd y\\
&=\lambda V_d R_c^d  \int_0^{1}  \exp{\left(
-\lambda V_d R_c^d y 
	\left(
		1+\int_1^{{\frac{1}{y}}}\frac{T}{T+t^{\frac{\alpha_0}{d}}
		}\intd t
	\right)
\right)} \intd y \iftoggle{SC}{}{ \\&\ \ \  }
+e^{-\lambda V_d R_c^d}.
\end{align*}
The second term goes to zero as $\lambda\rightarrow 0$. 
To prove the same for first term, consider an increasing sequence $\{\lambda_n\}$, and define
\iftoggle{SC}{$f_n(x)=\lambda_n \exp{\left(
-\lambda_n V_d R_c^d y 
	\left(
		1+\int_1^{{\frac{1}{y}}}\frac{T}{T+t^{\frac{\alpha_0}{d}}
		} \intd t
	\right)
\right)}.$}
{\begin{align*}
f_n(x)=\lambda_n \exp{\left(
-\lambda_n V_d R_c^d y 
	\left(
		1+\int_1^{{\frac{1}{y}}}\frac{T}{T+t^{\frac{\alpha_0}{d}}
		} \intd t
	\right)
\right)}.
\end{align*}}
It is clear that $f_n(x)\rightarrow 0$ pointwise for each $x$ in $(0,1)$. Also,
\iftoggle{SC}{$f_n(x)\le g(x)=\frac{1}{V_d R_c^de\left(
		1+\int_1^{{\frac{1}{y}}}\frac{T}{T+t^{\frac{\alpha_0}{d}}
		} \intd t
	\right)}.$}{\begin{align*}
f_n(x)\le g(x)=\frac{1}{V_d R_c^de\left(
		1+\int_1^{{\frac{1}{y}}}\frac{T}{T+t^{\frac{\alpha_0}{d}}
		} \intd t
	\right)}.
\end{align*}}
$g(x)$ is integrable on $(0,1)$ for $0\le \alpha_0 < d$. So by the dominance convergence theorem, first term, and hence the sum also goes to zero which proves the Theorem.

\bibliographystyle{IEEEtran}
\bibliography{references}

\end{document}